\PassOptionsToPackage{unicode}{hyperref}
\PassOptionsToPackage{hyphens}{url}
\PassOptionsToPackage{dvipsnames,svgnames,x11names}{xcolor}
\documentclass[12pt]{article}

\usepackage{amsmath,amssymb,mathtools}
\usepackage{iftex}
\ifPDFTeX
  \usepackage[T1]{fontenc}
  \usepackage[utf8]{inputenc}
  \usepackage{textcomp} 
\else 
  \usepackage{unicode-math}
  \defaultfontfeatures{Scale=MatchLowercase}
  \defaultfontfeatures[\rmfamily]{Ligatures=TeX,Scale=1}
\fi
\usepackage{lmodern}
\ifPDFTeX\else  
\fi
\IfFileExists{upquote.sty}{\usepackage{upquote}}{}
\IfFileExists{microtype.sty}{
  \usepackage[]{microtype}
  \UseMicrotypeSet[protrusion]{basicmath} 
}{}
\makeatletter
\@ifundefined{KOMAClassName}{
  \IfFileExists{parskip.sty}{%
    \usepackage{parskip}
  }{
    \setlength{\parindent}{0pt}
    \setlength{\parskip}{6pt plus 2pt minus 1pt}}
}{
  \KOMAoptions{parskip=half}}
\makeatother
\usepackage{xcolor}
\makeatletter
\ifx\paragraph\undefined\else
  \let\oldparagraph\paragraph
  \renewcommand{\paragraph}{
    \@ifstar
      \xxxParagraphStar
      \xxxParagraphNoStar
  }
  \newcommand{\xxxParagraphStar}[1]{\oldparagraph*{#1}\mbox{}}
  \newcommand{\xxxParagraphNoStar}[1]{\oldparagraph{#1}\mbox{}}
\fi
\ifx\subparagraph\undefined\else
  \let\oldsubparagraph\subparagraph
  \renewcommand{\subparagraph}{
    \@ifstar
      \xxxSubParagraphStar
      \xxxSubParagraphNoStar
  }
  \newcommand{\xxxSubParagraphStar}[1]{\oldsubparagraph*{#1}\mbox{}}
  \newcommand{\xxxSubParagraphNoStar}[1]{\oldsubparagraph{#1}\mbox{}}
\fi
\makeatother

\usepackage{longtable,booktabs,array}
\usepackage{calc} 
\usepackage{etoolbox}
\makeatletter
\patchcmd\longtable{\par}{\if@noskipsec\mbox{}\fi\par}{}{}
\makeatother
\IfFileExists{footnotehyper.sty}{\usepackage{footnotehyper}}{\usepackage{footnote}}
\makesavenoteenv{longtable}
\usepackage{graphicx}
\makeatletter
\def\maxwidth{\ifdim\Gin@nat@width>\linewidth\linewidth\else\Gin@nat@width\fi}
\def\maxheight{\ifdim\Gin@nat@height>\textheight\textheight\else\Gin@nat@height\fi}
\makeatother
\setkeys{Gin}{width=\maxwidth,height=\maxheight,keepaspectratio}
\makeatletter
\def\fps@figure{htbp}
\makeatother

\makeatletter
\@ifpackageloaded{caption}{}{\usepackage{caption}}
\AtBeginDocument{%
\ifdefined\contentsname
  \renewcommand*\contentsname{Table of contents}
\else
  \newcommand\contentsname{Table of contents}
\fi
\ifdefined\listfigurename
  \renewcommand*\listfigurename{List of Figures}
\else
  \newcommand\listfigurename{List of Figures}
\fi
\ifdefined\listtablename
  \renewcommand*\listtablename{List of Tables}
\else
  \newcommand\listtablename{List of Tables}
\fi
\ifdefined\figurename
  \renewcommand*\figurename{Figure}
\else
  \newcommand\figurename{Figure}
\fi
\ifdefined\tablename
  \renewcommand*\tablename{Table}
\else
  \newcommand\tablename{Table}
\fi
}
\@ifpackageloaded{float}{}{\usepackage{float}}
\floatstyle{ruled}
\@ifundefined{c@chapter}{\newfloat{codelisting}{h}{lop}}{\newfloat{codelisting}{h}{lop}[chapter]}
\floatname{codelisting}{Listing}

\makeatother
\makeatletter
\@ifpackageloaded{caption}{}{\usepackage{caption}}
\@ifpackageloaded{subcaption}{}{\usepackage{subcaption}}
\makeatother

\ifLuaTeX
  \usepackage{selnolig}  
\fi
\usepackage[]{natbib}
\usepackage{bookmark}

\IfFileExists{xurl.sty}{\usepackage{xurl}}{} 
\hypersetup{
  pdftitle={Discovering Spatial Patterns of Readmission Risk Using a Bayesian Competing Risks Model with Spatially Varying Coefficients},
  pdfauthor={Yueming Shen; Christian Pean; David Dunson; Samuel Berchuck},
  pdfkeywords={Geospatial analysis; Electronic health record data; Scalable Gaussian processes},
  colorlinks=true,
  linkcolor={blue},
  filecolor={Maroon},
  citecolor={Blue},
  urlcolor={Blue},
  pdfcreator={LaTeX via pandoc}}

\newcommand{\anon}{1}
\usepackage{bm}
\makeatletter
\@tfor\next:=abcdefghijklmnopqrstuvwxyzABCDEFGHIJKLMNOPQRSTUVWXYZ\do{%
  \def\command@factory#1{%
    \expandafter\def\csname B#1\endcsname{\mathbf{#1}}
  }
 \expandafter\command@factory\next
}

\@tfor\next:=ABCDEFGHIJKLMNOPQRSTUVWXYZ\do{%
  \def\command@factory#1{%
    \expandafter\def\csname mb#1\endcsname{\mathbb{#1}}
  }
 \expandafter\command@factory\next
}
\@tfor\next:=ABCDEFGHIJKLMNOPQRSTUVWXYZ\do{%
  \def\command@factory#1{%
    \expandafter\def\csname mc#1\endcsname{\mathcal{#1}}
  }
 \expandafter\command@factory\next
}
\makeatother
\newcommand{\cbr}[1]{\left(#1\right)}
\newcommand{\sbr}[1]{\left[#1\right]}
\DeclareMathOperator*{\argmin}{argmin}
\DeclarePairedDelimiter\abs{\lvert}{\rvert}

\begin{document}

\def\spacingset#1{\renewcommand{\baselinestretch}%
{#1}\small\normalsize} \spacingset{1}


\if1\anon
{
  \title{\bf Discovering Spatial Patterns of Readmission Risk Using a Bayesian Competing Risks Model with Spatially Varying Coefficients}
    \author{
    Yueming Shen\textsuperscript{1,2},
    Christian A. Pean\textsuperscript{3}, 
    David B. Dunson\textsuperscript{1}, and \\ Samuel I. Berchuck\textsuperscript{1,2,4} \\
    \textsuperscript{1}Department of Statistical Science, Duke University \\
    \textsuperscript{2}Duke AI Health, Duke University \\
    \textsuperscript{3} Department of Orthopaedic Surgery, Duke University \\
    \textsuperscript{4}Department of Biostatistics and Bioinformatics, Duke University
    }
  \maketitle
} \fi

\if0\anon
{
  \bigskip
  \bigskip
  \bigskip
  \begin{center}
    {\LARGE\bf Discovering Spatial Patterns of Readmission Risk Using a Bayesian Competing Risks Model with Spatially Varying Coefficients}
\end{center}
  \medskip
} \fi

\bigskip
\begin{abstract}

Time-to-event models are commonly used to study associations between risk factors and disease outcomes in the setting of electronic health records (EHR). In recent years, focus has intensified on social determinants of health, highlighting the need for methods that account for patients’ locations. We propose a Bayesian approach for introducing point-referenced spatial effects into a competing risks proportional hazards model. Our method leverages Gaussian process (GP) priors for spatially varying intercept and slope. To improve computational efficiency under a large number of spatial locations, we implemented a Hilbert space low-rank approximation of the GP. We modeled the baseline hazard curves as piecewise constant, and introduced a novel multiplicative gamma process prior to induce shrinkage and smoothing. A loss-based clustering method was then used on the spatial random effects to identify high-risk regions. We demonstrate the utility of this method through simulation and a real-world analysis of EHR data from Duke Hospital to study readmission risk of elderly patients with upper extremity fractures. Our results showed that the proposed method improved inference efficiency and provided valuable insights for downstream policy decisions.

\end{abstract}

\noindent%
\vfill

\newpage
\spacingset{1.8} 

\section{Introduction}

Hospital readmissions following surgical treatment for upper extremity fractures in older adults are a serious clinical and health system concern. These patients are often frail and medically complex, recovering from traumatic injuries that carry a high risk of complications, functional decline, and adverse events \citep{mathew2016risk, liu2022heavy, lee2025risk}. Consequently, they face both elevated readmission rates and substantial mortality risk, with many dying before a readmission can occur \citep{hao2019role, bourriquen2024effect, wang2024national}. Standard analyses of readmission outcomes that ignore this competing risk provide a misleading picture of patient prognosis and the factors driving readmission. At the same time, readmissions impose a heavy burden on patients and families and contribute major financial cost to the healthcare system, with average readmission costs 12.4\% higher than initial admissions in 2020 (\$16,300 vs. \$14,500) \citep{HCUP_readmissions_2023} and total readmission expenditures exceeding \$50 billion in 2018 \citep{weiss2021readmissions}. Together, these considerations make fracture readmissions an especially important setting for developing more sophisticated statistical tools. In particular, there is a need for methods that can accommodate the competing mortality risk while exploiting the rich patient-level information available in modern electronic health records (EHRs), including geocoded location data that capture unmeasured spatial determinants of health.

Recent work has recognized that patient outcomes are shaped not only by individual-level clinical characteristics but also by broader contextual factors, including geographic variation in healthcare access, socioeconomic conditions, and community-level resources \citep{vrtikapa2025social}. In the existing Bayesian competing risks literature, these unmeasured spatial determinants have been modeled using areal random effects, typically specified through conditional autoregressive (CAR) or intrinsic CAR (ICAR) priors. For example, CAR-based spatial frailty models have been applied to gastrointestinal cancer outcomes \citep{hesam2018spatial}, and ICAR extensions of the Fine-Gray model have been used in HIV/AIDS studies \citep{momenyan2022competing}. These approaches demonstrate that explicitly accounting for spatial dependence can improve inference on covariate effects and yield more accurate risk predictions when outcomes cluster geographically. However, areal models are designed for settings where patients are aggregated into regions such as counties, ZIP codes, or hospital catchments. In contrast, modern EHRs often contain patient-level geocoded addresses. Aggregating such data into arbitrary administrative units risks loss of resolution and residual confounding, motivating the need for point-referenced spatial models that directly leverage patient-level location information \citep{banerjee2003hierarchical}. To our knowledge, no existing work has integrated point-referenced spatial processes into a competing risks survival model.

In this article, we develop a Bayesian competing risks survival model that incorporates point-referenced spatial effects. This framework allows us to simultaneously account for the competing mortality risk that is central to older fracture patients, capture unmeasured spatial confounding at the patient level, and generate high-resolution maps of readmission risk. We use a novel multiplicative gamma process prior \citep{bhattacharya2011sparse} for the baseline hazard rates to encourage shrinkage and smoothing, and we model the spatial effects through Gaussian process (GP) priors. However, the scale of modern EHR data presents a serious computational barrier, as the number of unique patient locations can be very large. To address this, we adopt a Hilbert space GP approximation \citep{solin2020hilbert} that enables scalable Bayesian inference while retaining the key advantages of GP-based spatial modeling. In addition, following \cite{palmer2025quantifying}, we construct Bayes-optimal clusters for the spatial random effects to identify high-risk areas. While our motivating application is hospital readmission following upper extremity fracture, the proposed approach is broadly applicable to other clinical contexts where competing risks and spatial dependence are present. The remainder of the article is organized as follows. Section \ref{sec:data} describes our data sample, EHR data from Duke University. Section \ref{sec:method} introduces the proposed model, and Section \ref{sec:sim} reports results from a simulation study evaluating performance and scalability. Section \ref{sec:case} presents the data analysis of fracture readmissions, and finally Section \ref{sec:discussion} concludes with implications, limitations, and directions for future research.

\section{EHR Data and Exploratory Analysis}\label{sec:data}

The past decade has seen widespread adoption of EHRs in the U.S. healthcare system. In 2008, only 7.6\% of hospitals had a basic EHR system; by 2020, this number had risen to 81.2\% \citep{jiang2023pre}. EHRs now serve as indispensable infrastructure for clinical and population health research, offering large-scale, longitudinal, and granular patient-level data. Unlike registry or survey data, EHRs are not designed for research but are by-products of routine clinical care, which makes them both uniquely rich and methodologically challenging: They integrate diverse information sources (administrative, clinical notes, billing, imaging, lab results), follow patients irregularly through time, and frequently contain missing or inconsistent values. For this study, we sourced data from the Duke Enterprise Data Unified Content Explorer. It is a query platform that supports data exploration, cohort identification, and data extraction from Duke's enterprise data warehouse which contains raw EHR data with encounter records from three Duke-affiliated hospitals and more than 300 outpatient clinics \citep{horvath2014modular,hurst2021development}.

Our study included patients who underwent operative orthopaedic trauma fracture fixation of the upper extremity between January 2015 and March 2024, were at least 50 years old at the time of their initial fracture surgery, and resided in Durham county (where the main Duke hospital is located) at the time of the surgical procedure or one of the five neighboring North Carolina counties (Chatham, Orange, Person, Granville, or Wake). Patients were identified using Current Procedural Terminology (CPT) codes, aligned with Accreditation Council for Graduate Medical Education guidelines for orthopaedic trauma cases, together with International Classification of Diseases (ICD) diagnosis codes. Those receiving nonoperative management or presenting with multiple fractures were excluded. Full inclusion and exclusion CPT and ICD codes are documented in Section 1 of the Supplementary Materials. This study was approved by the Duke University Institutional Review Board and adhered to the Declaration of Helsinki and all applicable laws. It was approved with a waiver of informed consent due to the retrospective nature of the work. And because the underlying EHRs contain protected health information, we cannot share the data publicly.

\begin{figure}[t!]
  \centering
  \includegraphics[width=1\textwidth]{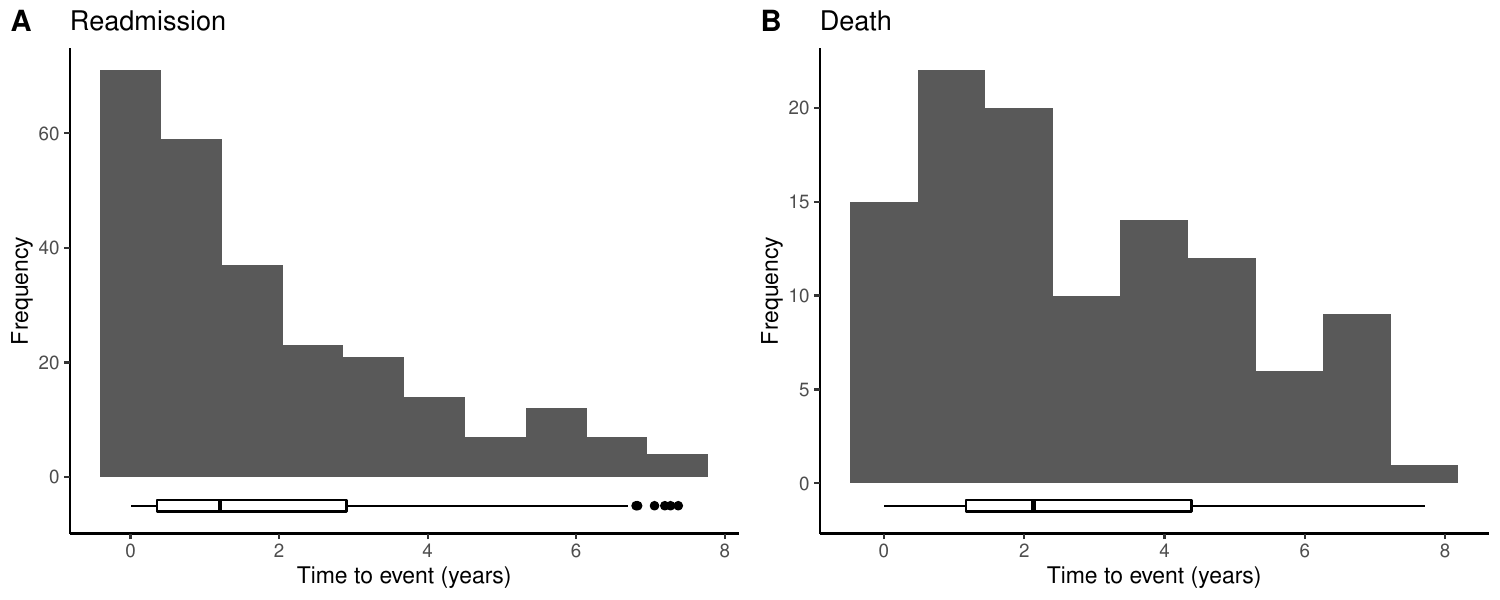}
  \caption{Distributions of the observed time to events. Panel A: Readmission events. Panel B: Death events.}
  \label{fig:EDA_hist_pop}
\end{figure}

A total of 1,245 patients met the criteria, of whom 43 had incomplete covariate information. Given the low missingness percentage, we performed a complete-case analysis on 1,202 patients. Among them, 255 experienced readmission and 109 died during the study period. See Figure \ref{fig:EDA_hist_pop} for distributions of the time to events. We see a spike of readmission within the first year of the initial surgery which is consistent with findings in existing clinical literature that readmissions are often front-loaded \citep{tian2023incidence}. Death events are comparatively more spread out over the 10 year period.

\begin{figure}[t!]
  \centering
  \includegraphics[width=14cm]{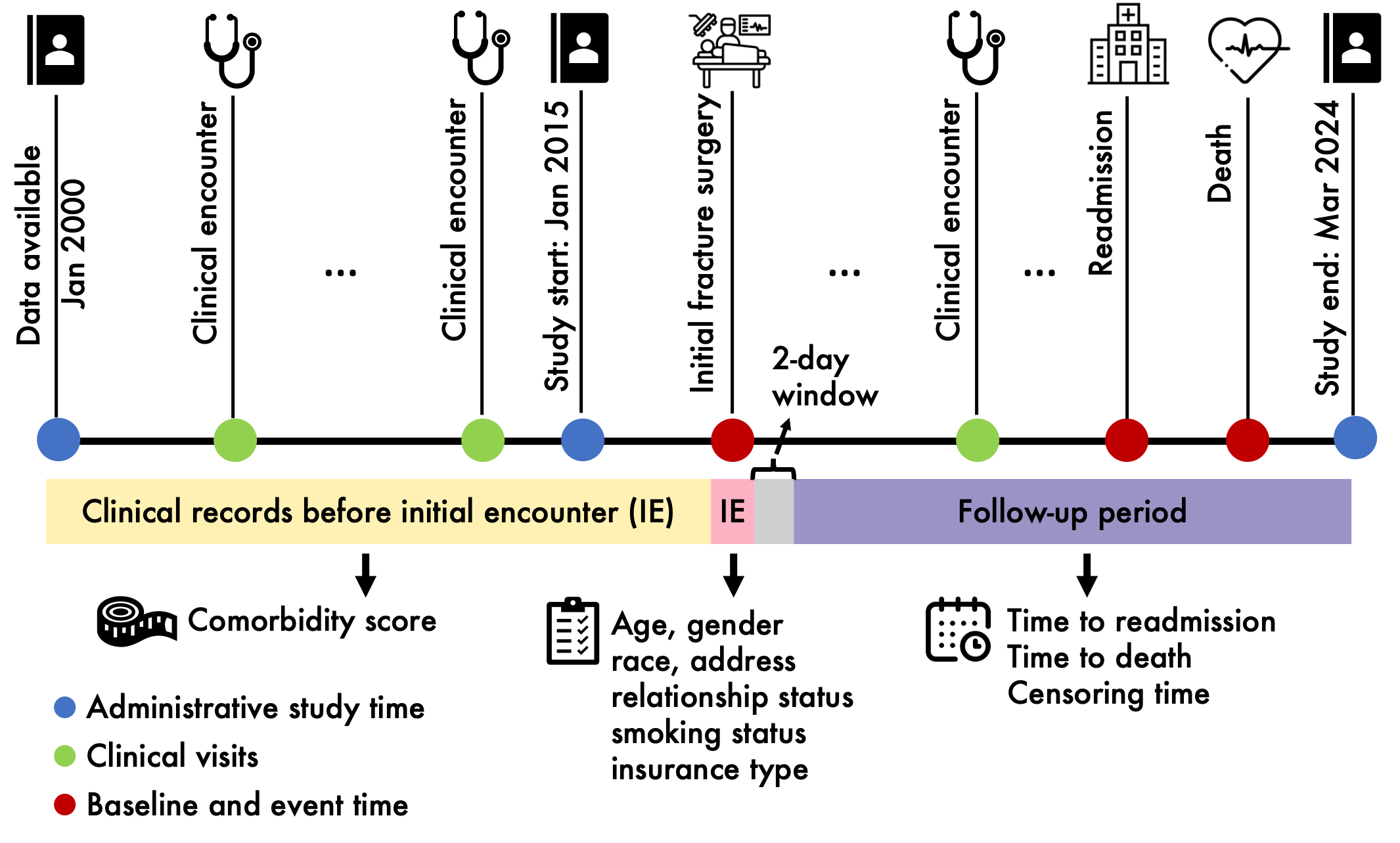}
  \caption{Visualizing the EHR data. Both the covariates and outcomes are defined based on the initial encounter (IE) date as an anchor. The comorbidity score is defined using medical records collected prior to the IE (time period in yellow); all other covariates are extracted at IE (time point in pink); event times are derived using clinical records at least two days after the IE (time period in purple). In this example, the patient was readmitted after the IE and then died, hence the observed data are readmission time $T$, with event type $\Delta=1$.}
  \label{fig:EHR}
\end{figure}

The EHRs capture a timeline of patient encounters across Duke–affiliated institutions (see Figure \ref{fig:EHR}). We extracted the following patient information at initial encounter: age, sex, race, smoking status, relationship status, insurance type, and residential location. See Section 3 of the Supplementary Materials for how we grouped insurance types from the raw EHR data. Medical records prior to the initial surgery were used to derive the Elixhauser readmission comorbidity score, following the 2024 Elixhauser comorbidity software user guide \citep{ahrq2024elixhauser}. This comorbidity score measures patient pre-existing health conditions, with higher values indicating higher readmission risk. Medical records two days after the initial surgery were used to extract time to events (see Figure \ref{fig:EHR}). The two-day window was used to ensure that events were separate from the initial encounters. Events of interest were death and readmission, with readmission defined as unplanned inpatient hospitalization, excluding psychiatric, hospice, rehabilitation, and elective admissions. Each patient was observed until the first occurrence of (1) readmission, (2) death, or (3) administrative censoring at the end of the study, i.e., for individual $i$, we observe  $(T_i, \Delta_i)$ where $T_i$ is time to first event and $\Delta_i \in \{1 \text{ (readmission)}, 2 \text{ (death)}, 0 \text{ (censored)}\}$. This induces a competing risks data structure: Patients who die cannot be subsequently readmitted, and readmitted patients are not followed further for mortality.

Table 1 in the Supplementary Materials presents the basic demographics of our patient cohorts, and we see notable differences in readmission rates across patient subgroups: Patients without partners had substantially higher rates (28\% vs. 17\%), as did ever-smokers (24\% vs. 19\%). By race, Black patients had the highest rate (28\%), compared with 21\% among White patients and 7\% among other minority groups. Patients with government insurance had the highest readmission rate at 28\%, with 13\% for commercial insurance, 11\% for WCSC, and 9\% for uninsured. This pattern likely reflects the fact that government-insured populations are generally older and have lower-income \citep{kff2024medicarecoverage,medpac2024databook}. As expected, age and comorbidity were positively associated with readmission risk: Readmitted patients were, on average, four years older than those not readmitted, and had comorbidity scores higher by 4.7 points. No gender differences were observed though, with approximately 21\% readmission rate in both men and women.

\begin{figure}[t!]
  \centering
  \includegraphics[width=1\textwidth]{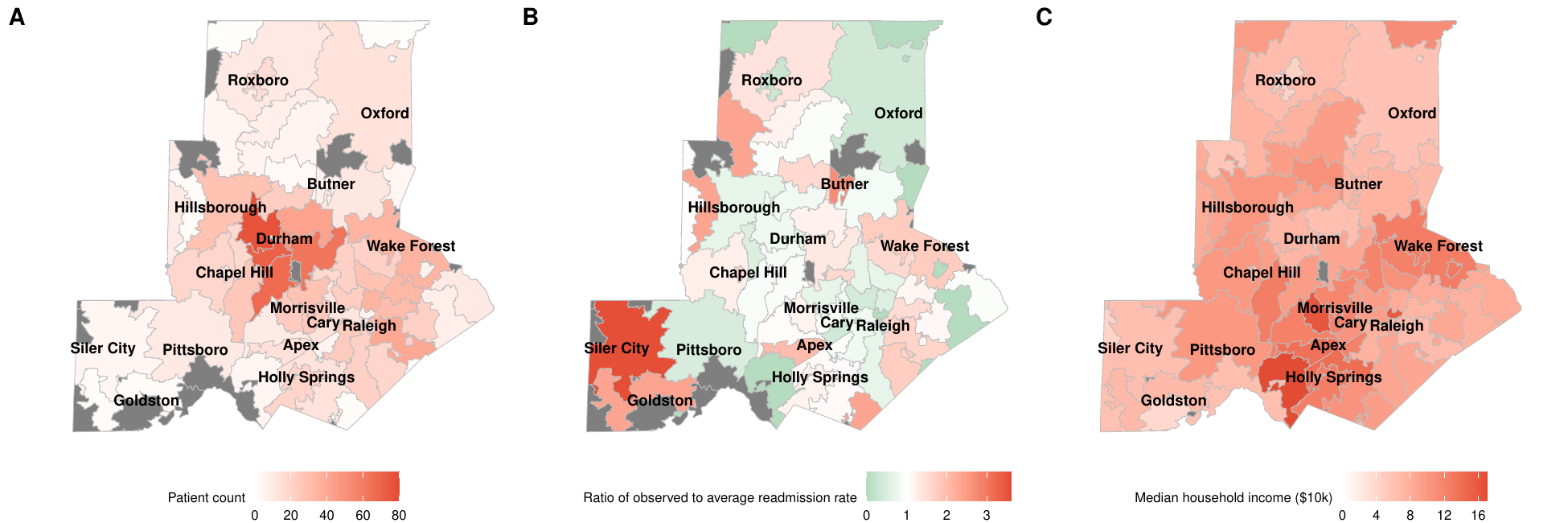}
  \caption{Spatial patterns of the empirical data at ZCTA level. Panel A: Patient count (gray areas indicate no patients). Panel B: Ratio of observed to average readmission rate. Panel C: 2023 median household income (in \$10k).}
  \label{fig:EDA_maps}
\end{figure}

Figure \ref{fig:EDA_maps} displays readmission outcomes aggregated to the Zip Code Tabulation Area (ZCTA) level in comparison to 2023 median household income data. Despite noise, spatial patterns emerge: Readmission rates are slightly elevated in East Durham, a predominantly lower-income area, and reduced in higher-income areas such as Chapel Hill, Morrisville and Cary. These patterns suggest potential spatial correlation in outcomes driven by unmeasured confounders, which motivate incorporation of spatial random effects.

\section{Model}\label{sec:method}

We propose a Bayesian competing risks proportional hazard model with spatially varying coefficients for point-referenced data. Section \ref{sec:crm} introduces the Bayesian competing risks model with proportional cause-specific hazard rates. Section \ref{sec:bhr} discusses how we model the baseline hazard rates as piecewise constant, and the use of a multiplicative gamma process prior for shrinkage. Section \ref{sec:spatial} presents the incorporation of spatial intercepts and spatial slopes through Gaussian process (GP) priors to capture point-referenced spatial dependence. Because full GP models are computationally prohibitive, Section \ref{sec:scalable} outlines a low-rank approximation method for scalability. Finally in Section \ref{sec:clustering}, we discuss a loss-based Bayes-optimal clustering method with uncertainty quantification.

\subsection{Bayesian Competing Risks Proportional Hazard Model}\label{sec:crm}

The two most commonly used competing risks regression models are the proportional sub-distributional hazard model, also known as the Fine-Gray model, and the proportional cause-specific hazard model \citep{haller2013applying}. The Fine-Gray model is suited for evaluating covariate effects on sub-distributional hazard which allows for influence from competing risk events, whereas the proportional cause-specific hazard model is preferable if the focus is etiological \citep{dignam2012use}. As we are primarily interested in covariate effects on readmission risk alone, we adopted the proportional cause-specific hazard model.

Suppose there are $n$ observations and $m$ competing risk events. Let $i=1,\dots,n$ index the observations, and $j=1,\dots,m$ index the risk types. For observation $i$, let $\Bx_i \in \mbR^p$ denote its covariates, $T_{i1}, \dots, T_{im}$ denote its latent event times, and $T_i = \min_{j \le m} T_{ij}$ denote time to the first event. The cause-specific hazard rate for risk type $j$ at time $t$ given covariates $\Bx_i$ is defined as $\lambda_j(t \mid \Bx_i) = \lim_{h \to 0} Pr[t \le T_i \le t+h, T_i=T_{ij} \mid T_i \ge t, \Bx_i]/h$. Under the proportional hazard assumption, $\lambda_j(t \mid \Bx_i) = \lambda_{0j}(t) \exp(\Bx_i^\top \bm{\beta}_j)$, where $\lambda_{0j}(t)$ is the baseline hazard rate for risk type $j$ at time $t$, and $\bm{\beta}_j \in \mbR^p$ are the regression coefficients for risk type $j$. The total hazard rate is the sum of the cause-specific hazard rates $\lambda(t \mid \Bx_i) = \sum_{j=1}^m \lambda_j(t \mid \Bx_i)$, and the survivor function is $S(t \mid \Bx_i) = \exp \cbr{- \int_0^t \lambda(u \mid \Bx_i)du}$. 

In this formulation, censoring is assumed to be independent of the event times, which is plausible for our application as censoring occurs only at the administrative study end. No intercept term is modeled because it is not separately identifiable from the baseline hazard level. To incorporate point-referenced spatial random effects, we implemented a full Bayesian analysis as frequentist methods would be computationally infeasible. For the regression coefficients $\bm{\beta}_j$'s, we used a normal inverse-gamma prior: $\bm{\beta}_j \sim N_p(\bm{0}, \sigma^2 \BI)$, $\sigma^2 \sim IG(a_\sigma,b_\sigma)$. Other priors are discussed in the following sections.

\subsection{Modeling the baseline hazard rates}\label{sec:bhr}

The baseline hazard curves are typically nuisance parameters. The simplest approach is to use parametric models, for example, assuming the event times follow Exponential or Weibull distributions \citep{wang2003bayesian, samanta2021bayesian}. Such parametric specifications yield parsimonious models but lack flexibility to capture complex hazard shapes in real data. A more flexible alternative is piecewise-constant hazard within pre-specified time intervals \citep{ibrahim2001bayesian}. Independent gamma priors on interval-specific hazard rates are commonly used \citep{hu2009bayesian}. This approach accommodates a wide range of hazard shapes, but without shrinkage across intervals, can produce unstable estimates in regions with sparse data; moreover, the prior is not coherent under changes to the time partition, requiring manual rescaling of hyperparameters for consistency. Nonparametric stochastic process priors address these issues by modeling the cumulative baseline hazard coherently and adaptively, without dependence on a fixed partition. Common choices include the gamma process prior \citep{kalbfleisch1978non} and the beta process prior \citep{hjort1990nonparametric} on the cumulative baseline hazard curve, both
leading to conjugate posterior distributions and 
inducing shrinkage towards a user-specified prior mean cumulative hazard. However, this shrinkage acts globally rather than locally, offering no direct control over smoothness of the hazard curves. Specifying a realistic prior mean function can also be difficult when prior information is limited.

To balance flexibility and computational tractability, we model the baseline hazard as piecewise constant with a multiplicative gamma process prior \citep{bhattacharya2011sparse}. Specifically, we partition time into $k$ intervals $0=s_0 < s_1 < \dots < s_k$, with $s_k$ set to the maximum observed event or censoring time. For risk type $j$, we let $\lambda_{0j}(t) = \lambda_{jl}$ for $t \in (s_l, s_{l+1}]$, and assign the following prior:
\begin{align*}
    \lambda_{jl} = \prod_{r=0}^l \psi_{jr}, \quad \psi_{j0} \sim Ga(a_0,b_0), \quad \psi_{jr \ge 1} \sim Ga \cbr{\frac{\kappa}{s_r - s_{r-1}},\frac{\kappa}{s_r - s_{r-1}}}, \quad \kappa \sim IG(a_1,b_1).
\end{align*}
This construction centers $\lambda_{jl}$ at $\lambda_{j(l-1)}$, inducing autocorrelation and hence shrinkage across adjacent intervals. The scaling by $(s_r - s_{r-1})^{-1}$ ensures that the degree of autocorrelation is invariant to the grid resolution. For fixed time points $t_1 < t_2$, the prior correlation between the corresponding baseline hazard rates $\lambda_{jl}$ and $\lambda_{j(l+q)}$ given $\kappa$, $a_0$ and $b_0$ is 
\begin{align*}
     \sbr{\frac{(1+a_0) \prod_{r=1}^l \cbr{(s_r-s_{r-1})/\kappa+1}-a_0}{(1+a_0) \prod_{r=1}^{l+q} \cbr{(s_r-s_{r-1})/\kappa+1}-a_0}}^{1/2},
\end{align*}
and it converges to $[((1+a_0)\exp(t_1/\kappa)-a_0)/((1+a_0)\exp(t_2/\kappa)-a_0)]^{1/2}$ as $k \to \infty$. The correlation increases as $a_0$ decreases or $\kappa$ increases, with $\kappa$ being the key driver. A hyperprior on $\kappa$ allows the data to inform the smoothness level.

This setup combines computational efficiency with adaptive smoothness. The piecewise-constant representation supports scalable inference, and with sufficiently many intervals, it can approximate a wide variety of hazard shapes; meanwhile, the multiplicative gamma process prior introduces coherent shrinkage across intervals. Together they yield stable, yet flexible, estimation of the baseline hazard curves.

\subsection{Modeling point-referenced spatial effects}\label{sec:spatial}

To introduce spatial effects into this model, letting $\Bd_i \in \mbR^2$ denote the geographic coordinates of observation $i$, we set 
\begin{align}\label{eq:crsm}
    \lambda_j(t_i \mid \Bx_i) = \lambda_{0j}(t_i) \exp\left(\Bx_i^\top \bm{\beta}_j + \theta_{0j}(\Bd_i) +(\theta_{1j}(\Bd_i) + \beta_{wj})w_i\right),
\end{align}
where $\theta_{0j}(\Bd_i)$ is the spatial intercept at location $\Bd_i$ for risk type $j$, $\theta_{1j}(\Bd_i)$ is the spatial slope, and $w_i \in \mbR$ is the covariate for which we study the spatial variation of its association with the risk events. Here we assume $w_i$ is not part of $\Bx_i$ and its fixed effect is captured by $\beta_{wj}$. In this model, level of the spatial intercepts is not separately identifiable from the level of the baseline hazard rates, and level of the spatial slopes is not separately identifiable from $\beta_{wj}$'s. We model point-referenced spatial dependence through GP priors:
\begin{align}\label{eq:GP}
    \theta_{0j}(\cdot) \sim GP\left(\bm{0}, K_\nu(\tau_{0j},\ell_{0j}) \right) \perp \theta_{1j}(\cdot) \sim GP\left(\bm{0}, K_\nu(\tau_{1j},\ell_{1j}) \right), 
\end{align}
where $K_\nu(\tau,\ell)$ denotes the Matérn covariance function with parameter $\nu$, magnitude parameter $\tau$ and lengthscale parameter $\ell$. We use independent truncated normal priors $TN(0,\sigma_\tau^2)$ for $\tau_{0j}$ and $\tau_{1j}$'s, and independent $Ga(a_\ell, b_\ell)$ priors for $\ell_{0j}$ and $\ell_{1j}$'s. Realistically, $\theta_{0j}$ and $\theta_{1j}$ are likely to be correlated, but we use independent priors because this dependence is not central to our objectives, is difficult to identify in a competing risks setting, and the independence assumption substantially improves computational efficiency.

An additional advantage of GP priors is that they facilitate kriging, i.e., spatial prediction at new, unobserved locations \citep{banerjee2003hierarchical}. Taking the spatial intercept as an example. Let $\bm{\theta}_{0j} \in \mbR^n$ denote the spatial intercepts for risk type $j$ at observed locations, and suppose we are interested in the spatial intercepts for $q$ new locations $\Bd^*_1, \dots \Bd^*_q$, denoted as $\bm{\theta}^*_{0j} \in \mbR^q$. Let $(\BT, \bm{\delta})$ denote the data where $\BT \in \mbR^n$ are the event times, and $\bm{\delta} \in \{0,1,2\}^n$ are the event types. Let $\bm{\Omega}=(\bm{\beta},\bm{\lambda}, \bm{\tau}, \bm{\ell})$ denote all the model parameters consisting of the regression coefficients, baseline hazard rates, GP magnitude and lengthscale parameters. The posterior predictive distribution for $\bm{\theta}^*_{0j}$ can be obtained as:
\begin{align*}
    f(\bm{\theta}^*_{0j} \mid \BT, \bm{\delta}) = \int f(\bm{\theta}^*_{0j} \mid \bm{\theta}_{0j}, \bm{\Omega}) f(\bm{\theta}_{0j}, \bm{\Omega} \mid \BT, \bm{\delta}) d \bm{\theta}_{0j} d\bm{\Omega},
\end{align*}
where $f(\bm{\theta}_{0j}, \bm{\Omega} \mid \BT, \bm{\delta})$ is the posterior distribution for $\bm{\theta}_{0j}$ and $\bm{\Omega}$, and $f(\bm{\theta}^*_{0j} \mid \bm{\theta}_{0j}, \bm{\Omega})$ is the kriging distribution. Given $\bm{\Omega}$, let $\bm{\Sigma} \in \mbR^{n \times n}$ denote the covariance matrix for $\bm{\theta}_{0j}$, $\bm{\Sigma}^* \in \mbR^{q \times q}$ denote the covariance matrix for $\bm{\theta}_{0j}^*$, and $\bm{\Sigma}_+ \in \mbR^{n \times q}$ denote their cross covariance matrix. Under the GP, $\bm{\theta}_{0j}$ and $\bm{\theta}_{0j}^*$ are jointly multivariate normal given $\bm{\Omega}$:
\begin{align*}
    \begin{pmatrix}
        \bm{\theta}_{0j} \\
        \bm{\theta}_{0j}^*
    \end{pmatrix} \Bigg| \bm{\Omega} \sim N\cbr{
    \begin{pmatrix}
        \bm{0}_n \\
        \bm{0}_q
    \end{pmatrix},
    \begin{pmatrix}
        \bm{\Sigma} & \bm{\Sigma}_{+} \\
        \bm{\Sigma}_+^\top & \bm{\Sigma}^*
    \end{pmatrix}}. 
\end{align*}
Therefore the kriging distribution is a $q$-dimensional multivariate normal by standard properties of the multivariate normal distribution:
\begin{align}\label{eq:kriging}
    \bm{\theta}_{0j}^* \mid \bm{\theta}_{0j}, \bm{\Omega} \sim N(\bm{\Sigma}_+^\top \bm{\Sigma}^{-1}\bm{\theta}_{0j}, \bm{\Sigma}^*-\bm{\Sigma}_+^\top\bm{\Sigma}^{-1}\bm{\Sigma}_+).
\end{align}
This closed-form expression enables easy posterior predictive sampling of spatial random effects at kriging locations within the Bayesian model fitting procedure, providing predictions with uncertainty quantification.

\subsection{Scalable Gaussian Process}\label{sec:scalable}

Modeling spatial intercepts and spatial slopes for the two competing risk events in our data requires four GPs. A full GP implementation scales as $\mcO(n^3)$ in computation and $\mcO(n^2)$ in memory, which is computationally infeasible for our data even with a moderate sample size of $n=1,202$. Therefore, we explored scalable methods and implemented Hilbert Space GP (HSGP), a low-rank approximation method introduced by \cite{solin2020hilbert}. HSGP approximates the $n \times n$ covariance matrix $\bm{\Sigma}$ as $\bm{\Phi} \BS \bm{\Phi}^\top$, where $\bm{\Phi} \in \mbR^{n \times m}$ is a feature matrix which only depends on the observed locations and the size of the geographical area for approximation, $\BS \in \mbR^{m \times m}$ is diagonal which only depends on the covariance function and its parameters, and $m$ is the number of basis functions. This approximation reduces the computation complexity from $\mcO(n^3)$ to $\mcO(nm+m)$. 

Kriging under HSGP is also straightforward. Let $\bm{\Phi}^* \in \mbR^{q \times m}$ denote the feature matrix for the kriging locations. If $m \ge n$, the kriging distribution is analogous to equation \eqref{eq:kriging}:
\begin{align*}
    \bm{\theta}_{0j}^* \mid \bm{\theta}_{0j}, \bm{\Omega} \sim N( (\bm{\Phi}^*\BS\bm{\Phi}^\top)(\bm{\Phi}\BS\bm{\Phi}^\top)^{-1}\bm{\theta},(\bm{\Phi}^*\BS\bm{\Phi}^{*\top})-(\bm{\Phi}^*\BS\bm{\Phi}^\top)(\bm{\Phi}\BS\bm{\Phi}^\top)^{-1}(\bm{\Phi}\BS\bm{\Phi}^{*\top})).
\end{align*}
If $m<n$, $(\bm{\Phi}\BS\bm{\Phi}^\top)$ is not invertible. In this case the kriging distribution is degenerate, with $\bm{\theta}_{0j}^* \mid \bm{\theta}_{0j}, \bm{\Omega} = (\bm{\Phi}^*\BS\bm{\Phi}^\top)(\bm{\Phi}\BS\bm{\Phi}^\top)^+ \bm{\theta}_{0j}$, where $\BA^+$ denote a generalized inverse of matrix $\BA$. In particular, when using a non-centered parameterization in \texttt{Stan} \citep{stan2025} where $\bm{\theta}_{0j}$ is sampled as $\bm{\Phi} \BS^{1/2} \Bz$ for $\Bz \sim N_m(\bm{0},\BI)$, kriging can be done by simply setting $\bm{\theta}_{0j}^* \mid \bm{\theta}_{0j}, \bm{\Omega}$ to $\bm{\Phi}^*\ \BS^{1/2}\Bz$. See Section 4 of the Supplementary Materials for details.

We followed \cite{riutort2023practical} to implement HSGP, and used an iterative procedure to decide $m$ and the other tuning parameters. We also carried out a simulation study to verify HSGP approximation accuracy. See discussions in Section \ref{sec:sim}.

\subsection{Loss-based clustering}\label{sec:clustering}

One of our research goals is to identify high-risk areas. \cite{palmer2025quantifying} proposed a Bayesian decision theoretic approach to cluster patients in non-spatial settings based on the posterior distribution of their random effects and a loss function. We adapt this approach to spatial clustering, producing point estimates and uncertainty quantification for clusters.

For each spatial process, we want to estimate a partition $\BC=\{C_1,\dots,C_K\}$ for the kriging locations, where each $C_k \subset \{1,\dots,q\}$ are indices such that the locations are placed into $K$ distinct clusters. To simplify the notation, let $\theta_1, \dots, \theta_q$ denote the spatial random effects at $q$ kriging locations, and let $\pi_{\text{post}}$ denote their posterior distribution. We seek cluster labels $\Bl=(l_1,\dots,l_q)$ and cluster centers $\Bc=(c_1,\dots,c_K)$ which minimize the posterior expected K-means loss: $(\hat{\Bl},\hat{\Bc}) = \argmin_{\Bl:\abs{\BC}=K, \Bc \in \mbR^K} \sum_{k=1}^K \sum_{i \in C_k} E_{\pi_{\text{post}}}[(\theta_i - c_k)^2]$. As shown by Proposition 3.1 in \cite{palmer2025quantifying}, this is equivalent to solving for:
\begin{align*}
    (\hat{\Bl},\hat{\Bc}) = \argmin_{\Bl:\abs{\BC}=K, \Bc \in \mbR^K} \sum_{k=1}^K \sum_{i \in C_k} (E_{\pi_{post}}[\theta_i] - c_k)^2,
\end{align*}
which can be easily obtained by applying K-mean clustering to the posterior mean of the $\theta_i$'s. Conditioning on the estimated cluster centers $\hat{\Bc}$, we can characterize uncertainty in clustering through the posterior probability of assigning each location $i$ to cluster $k$:
\begin{align*}
    Pr(l_i=k \mid \Bc=\hat{\Bc}) = E_{\pi_{\text{post}}} \sbr{1\cbr{ \argmin_{k' \in \{1, \dots, K\}} (\theta_i - \hat{c}_{k'})^2 = k}}.
\end{align*}

\section{Simulation study motivated by EHR data}\label{sec:sim}

Motivated by the EHR data application, we designed a simulation study with three objectives: (1) understand the performance of our proposed model under model misspecification; we used independent Gaussian process (GP) priors between spatial slopes and intercepts of the same risk event, and also across risk events, we wanted to see how this model performs under EHR data generated from more complex spatial dependence structure, (2) compare the proposed Bayesian model performance with frequentist competitors, and (3) evaluate the accuracy of the Hilbert Space GP (HSGP) approximation to the full GP. 

To better emulate real EHR data, we based the simulation study on a subset of our application data. The sample size was chosen to be $n=225$ so that full GP implementation was feasible. Prior to use, the data were de-identified by capping age at 80, jittering locations, after which the coordinates were centered and rescaled so that all observations lie within the square $[-1,1] \times [-1,1]$ centered at the origin. Using all available covariates and locations, we generated competing risks datasets from the model in equation \eqref{eq:crsm}. But instead of using independent spatial intercepts and slopes described in equation \eqref{eq:GP}, we simulated the spatial effects using the linear model of coregionalization (LMC) \citep{banerjee2003hierarchical}, where the spatial intercepts and slopes were generated as linear combinations of eight independent but not identically distributed GPs. This setup enabled assessment of our model’s performance for data with more complicated spatial patterns.

We generated 500 competing risks datasets with $m=2$ risk types, each with 40\% censored observations. There were a total of $p=11$ covariates, including age, sex, race, smoking status, relationship status, insurance type and comorbidity score, among which we modeled spatial slopes for comorbidity. The true regression coefficients for comorbidity were set to $0.5$ for risk type 1 and $-0.7$ for risk type 2. The other true regression coefficient $\bm{\beta}_j$'s were generated from $N(0,0.25)$ and then fixed for all the datasets. The true baseline hazard rates were parameterized as $\lambda_{0j}(t) = \gamma_j \alpha_j t^{\alpha_j-1} \exp(c_j)$, with $\gamma_1 = 1$, $\gamma_2 = 2$, $a_1 = 10/3$, $\alpha_2=2$, $c_1 = -10$, and $c_2 = 5$. We used $k=30$ equally distanced knots, with $s_k$ set to the maximum event/ censoring time of each simulated dataset. The remaining hyperparameters matched those in the real-data analysis: $a_\sigma=1$, $b_\sigma=1$, $a_0=1$, $b_0=1$, $a_1=2$ and $b_1=40$. Matérn $3/2$ covariance function was used for the GPs, with $\sigma_\tau^2=16$, $a_\ell=2$, and $b_\ell=1$ as its hyperparameters. Section \ref{sec:case} provides the rationale for these choices.

We carried out Bayesian analysis using both GP and HSGP, where four parallel chains were run, each with 1,000 burn-in, followed by 4,000 iterations, thinned every 5th iteration. In order to study the spatial pattern, we set the kriging locations to be on a $21 \times 21$ grid in the $2 \times 2$ square, and obtained posterior predictive samples for $\bm{\theta}^*_{0j}$ and $\bm{\theta}^*_{1j}$'s at these locations. 

For comparison, we considered two frequentist models. The first (Coxph) was a competing risks proportional hazards model without spatial effects, implemented via the \texttt{coxph} function in the \texttt{survival} R package \citep{survival-package}. The same function was used for the second model (Coxph+group) where we further grouped individuals into 9 location-based clusters using the R \texttt{kmeans} function, and included cluster label as a covariate to approximate areal spatial effects. And we made sure all the runs converged within the maximum iteration limit. We chose to use Coxph+group to capture spatial effects because we were unable to identify a reliable implementation of competing risks models with random effects in standard R packages.

Now we present model performance results on the regression coefficients. Table \ref{tb:sim_beta1} reports root mean squared errors (RMSE) of the regression coefficients for risk type 1, computed using the true parameters and point estimates from the 500 simulated datasets. For the frequentist methods, point estimates were taken directly from the fitted models, while for the Bayesian models, we used the posterior means. As the fixed effect for comorbidity ($\beta_{w,1}$) is not separately identifiable from the level of the spatial slopes, we constrained the posterior spatial slopes across all kriging locations to sum to zero. Under this constraint, posterior estimates of $\beta_{w,1}$ represented the fixed effect and were directly comparable to the frequentist estimates. The first two rows of Table \ref{tb:sim_beta1} are results from the frequentist models, followed by the Bayesian spatial model with a full GP (BSp GP) in row 3 and its HSGP approximation (BSp HSGP) in row 4. BSp GP and BSp HSGP produced nearly identical results, demonstrating that HSGP offers a good approximation for GP under our model. Although the data were generated under the more complex LMC structure than the assumed model, the Bayesian spatial models consistently outperformed the frequentist competitors. This improvement was driven by two factors: richer spatial modeling, and shrinkage in the Bayesian framework which reduces posterior variance. The shrinkage effect was especially evident for covariates with limited information. For instance, $\beta_{8,1}$ and $\beta_{9,1}$ correspond to Insurance–WCSC and Insurance–Selfpay, where the subset of data we used for the simulation included only two readmission cases with WCSC and none with Selfpay. Consequently, the corresponding frequentist estimates exhibit high variance and hence much larger RMSEs than their Bayesian counterparts. The only exception was $\beta_{w,1}$, for which the Bayesian and frequentist models perform similarly. This was expected because incorporating random slopes increased the posterior uncertainty. Overall, these findings demonstrate the advantages of the proposed Bayesian model over existing frequentist alternatives. Results for risk type 2 are similar and can be found in Section 5 of the Supplementary Materials.

{%
\renewcommand{\arraystretch}{0.6}
\begin{table}[t!]
\centering
\caption{Risk type 1 regression coefficients RMSE from simulation study. \label{tb:sim_beta1}}
{\footnotesize
\begin{tabular}{l r r r r r r r r r r r}
\toprule
  \textbf{Model} &$\beta_{1,1}$ & $\beta_{2,1}$ & $\beta_{3,1}$ & $\beta_{4,1}$ & $\beta_{5,1}$ & $\beta_{6,1}$ & $\beta_{7,1}$ & $\beta_{8,1}$ & $\beta_{9,1}$ & $\beta_{10,1}$ & $\beta_{w,1}$ \\
\midrule
Coxph & 1.33 & 3.75 & 0.72 & 0.41 & 0.84 & 0.44 & 0.53 & 2.65 & 10.78 & 0.26 & 0.37\\
Coxph+groups & 1.51 & 3.87 & 0.77 & 0.50 & 1.04 & 0.49 & 0.62 & 2.74 & 10.93 & 0.31 & 0.39\\
BSp GP & 0.31 & 0.24 & 0.25 & 0.31 & 0.30 & 0.24 & 0.29 & 0.44 & 0.51 & 0.23 & 0.38\\
BSp HSGP & 0.31 & 0.24 & 0.25 & 0.31 & 0.30 & 0.24 & 0.29 & 0.45 & 0.51 & 0.23 & 0.39\\
\bottomrule
\end{tabular}
}
\end{table}
}

\begin{figure}[t!]
  \centering
  \includegraphics[width=1\textwidth]{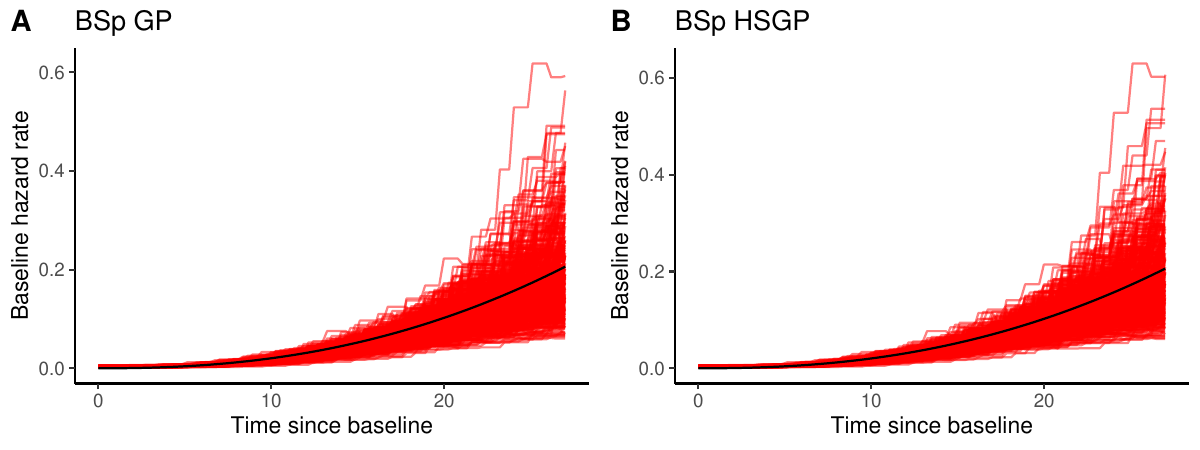}
  \caption{Risk type 1 baseline hazard rate results from simulation study with true hazard curves in black, and model fitted posterior means in red. Panel A: GP. Panel B: HSGP.}
  \label{fig:sim_r1_bhr}
\end{figure}

Next, we examine results for the baseline hazard rates and the spatial surfaces, which were only available from the Bayesian models. Figure \ref{fig:sim_r1_bhr} shows the posterior means of the piecewise-constant baseline hazard rates for risk type 1 from the simulated datasets, compared to the true curve. Although the true hazard curve was smooth rather than piecewise-constant, the model recovered them well, particularly in the early periods where more information was available. We further computed RMSE for each dataset by comparing the posterior samples with the truth (average of true rates at interval end points were used as benchmark). The RMSEs were all below 0.005 for the first five knots, rising to about 1 in later periods. Overall, these results showed that the proposed model can effectively learn the baseline hazard curves, and that GP and HSGP produce comparable estimates. See Section 5 of the Supplementary Materials for RMSE plots and results for risk type 2.

\begin{figure}[t!]
  \centering
  \includegraphics[width=1\textwidth]{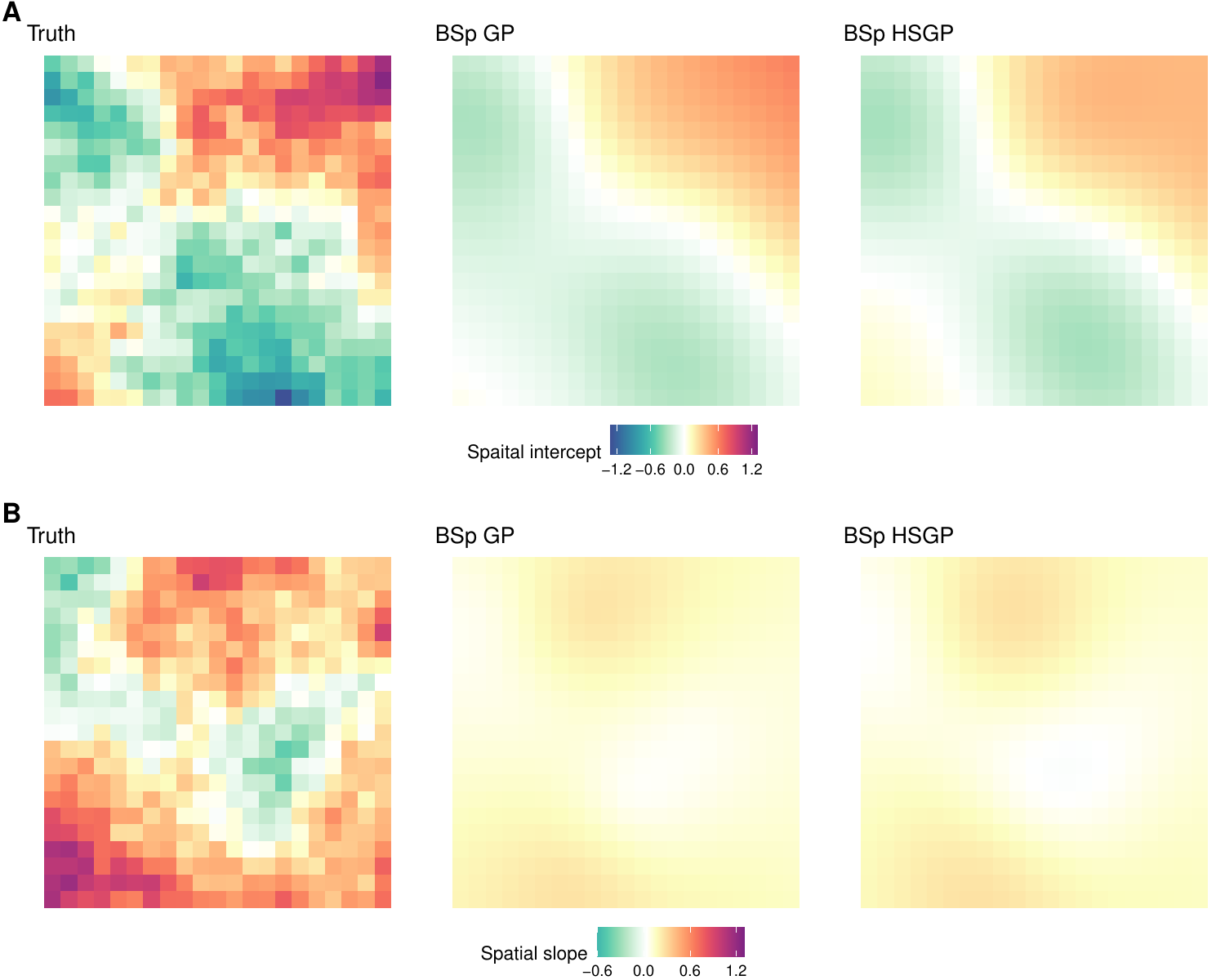}
  \caption{Risk type 1 spatial surface estimation results from simulation study. Panel A: the true spatial intercept surface, average of posterior means from GP and HSGP. Panel B: the true spatial slope surface, average of posterior means from GP and HSGP.\label{fig:sim_r1_sp_m}}
\end{figure}

\begin{figure}[t!]
  \centering
  \includegraphics[width=1\textwidth]{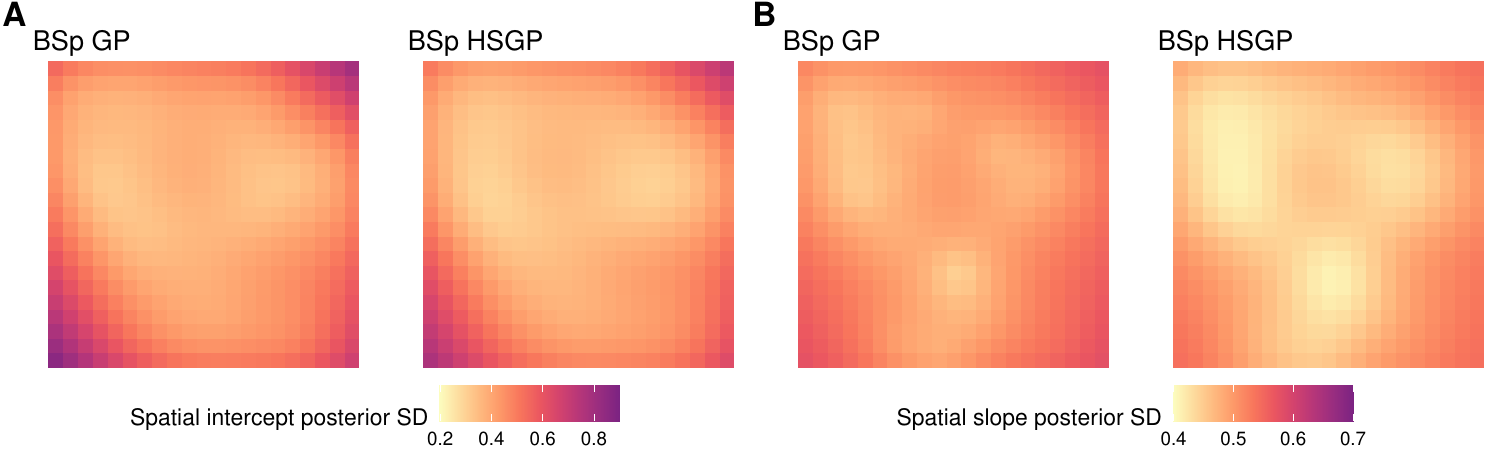}
  \caption{Average of posterior standard deviations for risk type 1 spatial surfaces from simulation study. Panel A: Spatial intercept results form GP and HSGP. Panel B: Spatial slope results from GP and HSGP.\label{fig:sim_r1_sp_sd}}
\end{figure}

Figure \ref{fig:sim_r1_sp_m} compares the true spatial surfaces with the averages of the posterior means for risk type 1, and panel A is for spatial intercepts. Because the levels of the spatial intercepts and the baseline hazard rates are not separately identifiable, we constrained the posterior spatial intercepts at all kriging locations to sum to zero, and applied the same adjustment to the true intercept surface for comparability. As an artifact of the LMC, the true intercept surface was not smooth. The model recovered the broad spatial pattern but did not capture finer local fluctuations. Panel B is for spatial slopes. Since the fixed effect and the level of the slopes are not separately identifiable, we presented results for their sum. Given the structure of the competing risks model (see discussions in Section \ref{sec:method_imp} on the weak identifiability of large slope values), slope surfaces are harder to estimate. Accordingly, the model only captured coarse patterns, with much of the fine-scale detail missing. Figure \ref{fig:sim_r1_sp_sd} presents uncertainty quantification results. Panel A shows the average posterior standard deviations for the spatial intercepts across simulated datasets, and panel B shows the same for the spatial slopes. For confidentiality reasons, we did not display the observation locations, but we observed a clear pattern that the posterior uncertainty is lower in regions with more observations and higher in regions with fewer or no observations. This pattern was typical of GP-based kriging and aligned with expectations. Additional uncertainty quantification results including RMSE and posterior predictive coverage of the spatial surfaces, as well as results for risk type 2 can be found in Section 5 of the Supplementary Materials. 

HSGP and GP produced broadly similar results, with only a few notable differences. In panel A of Figure \ref{fig:sim_r1_sp_m}, HSGP captured the high-risk area in the bottom-left corner slightly better, likely because it mixed more efficiently and achieved higher effective sample sizes than GP. In contrast, Figure \ref{fig:sim_r1_sp_sd} shows that HSGP tended to underestimate posterior uncertainty. This was expected since HSGP is a low-rank approximation method. For example, if the number of kriging locations exceeded the number of basis functions used in HSGP, the kriging distribution was necessarily degenerate, leading to underestimation of the posterior uncertainty. Despite this limitation, HSGP offered substantial computational gains. Across 500 runs, the full GP averaged 8.3 hours (2.5\% and 97.5\% quantiles: 3.8–13.8 hours), whereas HSGP averaged only 0.7 hours (2.5\% and 97.5\% quantiles: 0.2–1.6 hours). Thus, while yielding comparable inference, full GP was on average an order of magnitude slower than HSGP. And this gap was expected to widen further with increasing sample size.

Overall, the simulation results showed that although our model assumed a simplistic spatial dependence structure, it improved estimation efficiency for the regression coefficients even under model misspecification where the data exhibited more complex spatial dependence. The findings also highlighted the advantages of the proposed Bayesian approach over available frequentist alternatives and confirmed the accuracy of the HSGP approximation, making it a scalable option for our real data application.

\section{Application to EHR data}\label{sec:case}

We implemented three models to analyze the EHR data: a Bayesian competing risks model without spatial effects (BNSp), a model with spatial intercepts only (BSp i), and a model with both spatial intercepts and slopes (BSp i+s). Details of model implementation are given in Section \ref{sec:method_imp}, with posterior convergence checks and model comparison described in Section \ref{sec:method_com}. Section \ref{sec:results} presents the fitting and clustering results from the selected model, and Section \ref{sec:sensi} reports sensitivity analyses on the baseline hazard hyperparameters.

\subsection{Model Implementation}\label{sec:method_imp}

We adopted the same setup as in the simulation study for all three models: Four parallel chains, each with 1,000 burn-in, followed by 4,000 iterations, thinned every fifth iteration, yielding a total of 3,200 posterior samples. For the hyperparameters, we assigned a weakly informative $IG(a_\sigma=1,b_\sigma=1)$ prior to $\sigma^2$, and weakly informative $Ga(a_0=1,b_0=1)$ priors for the $\psi_{j0}$'s. In our application, the data span 480 weeks, and we set $k=50$ with equally spaced knots. We then used a $Ga(a_1=2,b_1=40)$ hyperprior for $\kappa$ as it allows $\kappa$ to vary over a range that induces plausible autocorrelation structures for the baseline hazard rates. In the absence of prior information about the smoothness of the hazard curves, we performed sensitivity analyses using different values of $a_0$, $b_0$, $a_1$, $b_1$ and $k$ to assess the robustness of our model results. See Section \ref{sec:sensi} for details.

We used the Matérn $3/2$ covariance function ($\nu=3/2$) for the Gaussian processes (GPs). It yields rougher sample paths than the Radial Basis Function kernel and is well suited for modeling processes expected to exhibit moderate, rather than high, smoothness. To specify priors for the GP hyperparameters, we considered the range of plausible parameter values. In competing risks models, large absolute values of regression coefficients are nearly non-identifiable: Under the proportional hazards assumption, an increase of the regression coefficient by 1 leads to $\exp(1) \approx 2.72$ times the hazard rate. As a result, it is hard for the model to distinguish between two large positive (negative) regression coefficient values because under either one, patients with positive (negative) covariates will fail almost instantaneously. With this in mind, we used an independent $\text{Half-Normal}(0,\sigma_\tau^2=16)$ prior for the magnitude parameters $\tau_{0j}$ and $\tau_{1j}$'s, which constrains the variance and thereby the values of the spatial effects to a reasonable range. Similar considerations guided the choice of priors for the lengthscale parameters. We wanted the prior to accommodate lengthscales close to zero to prevent over-smoothing the spatial surfaces, and we wanted its probability mass to concentrate on values that induce realistic correlation structures across the study region. Under the Matérn $3/2$ covariance function, a lengthscale of 1 corresponds to about 25\% correlation for locations 10 miles apart, with correlation tapering off by 30 miles, which is reasonable in practice. In contrast, a lengthscale of 5 implies 50\% correlation at 30 miles and 25\% at 50 miles, which is unrealistically high; thus, 5 serves as a practical upper bound. With these considerations, we placed an independent $IG(a_\ell=2,b_\ell=1)$ prior on $\ell_{0j}$ and $\ell_{1j}$'s, truncated above at 10. This prior has mean 1 and mode 1/3, with most of its probability mass within $(0,5)$. At the same time, its heavy tail permits larger lengthscale values if supported by the data.

Continuous covariates were scaled and centered, binary covariates were also centered because any constant in the regression term will not be separately identifiable from the level of the baseline hazard rates. Patient geocoordinates were transformed to plane coordinates using the Universal Transverse Mercator projection prior to spatial modeling. For kriging, we constructed a $37 \times 38$ grid over the study region; after excluding locations outside the six counties of interest, this produced 907 prediction sites. The spatial models were implemented using the Hilbert Space GP approximation.

\subsection{Model Checking and Model Comparison}\label{sec:method_com}

We assessed posterior convergence diagnostics for all three models and found no evidence of non-convergence. Effective sample sizes were approximately 3,000 for most parameters and exceeded 1,800 in all cases. Traceplots for selected parameters are provided in Section 6 of the Supplementary Material.

We used the Watanabe–Akaike information criterion (WAIC) \citep{watanabe2010asymptotic} for model comparison and selection. WAIC is well suited for complex Bayesian models because it is computationally efficient and asymptotically equivalent to a Bayesian leave-one-out cross validation \citep{watanabe2010asymptotic}. WAIC values were computed using the \texttt{loo} R package \citep{looR} based on pointwise log-likelihoods. Among the three models, BSp i+s achieved the lowest WAIC (4569.31), followed by BSp i (4573.92) and BNSp (4582.14). Since lower WAIC values indicate better performance, these results support including both spatial intercepts and spatial slopes in our analysis. 

\subsection{Associations and Geospatial Patterns of Readmission Risk}\label{sec:results}

In this section, we present results of the selected model BSp i+s. Table \ref{tb:res_r1_beta} reports posterior means and 95\% credible intervals of the regression coefficients on hazard ratio scale (i.e., $\exp(\bm{\beta_j})$). As intercepts were not identifiable, these results were relative to the following baseline patient profile: White male, never smoked, with partner, and with government insurance. Commercial insurance emerged as a protective factor for readmission risk, which may reflect the higher income levels of commercially insured patients \citep{call2022factors}. Having no partner was a significant factor for higher readmission risk, which is plausible considering the lack of support. As expected, both age and comorbidity scores were also significantly and positively associated with readmission risk. Overall, even after controlling for spatial confounders and other covariates, the model fitting results were quite similar to the EDA results.

{%
\renewcommand{\arraystretch}{0.6}
\begin{table}[t!]
\centering
\caption{Posterior means and 95\% credible intervals for $\bm{\beta}_j$'s on hazard ratio scale. \label{tb:res_r1_beta}}
{\footnotesize
\begin{tabular}[t]{l c c c c}
\toprule
\multicolumn{1}{l}{ } & \multicolumn{2}{c}{Readmission risk} & \multicolumn{2}{c}{Mortality risk} \\
Variable  & Mean & 95\% CI & Mean & 95\% CI \\
 \midrule
{Age} (per 1 SD) & 1.31 & (1.11, 1.54) & 1.68 & (1.19, 2.32)\\
{Comorbidity} (per 1 SD) & 1.90 & (1.49, 2.40) & 1.92 & (1.21, 2.93)\\
Female sex (baseline: Male) & 0.90 & (0.65, 1.20) & 0.93 & (0.50, 1.64)\\
\multicolumn{5}{l}{{Race} (baseline: White)} \\
\hspace{1em} Black & 1.34 & (0.90, 1.89) & 1.17 & (0.54, 2.23)\\
\hspace{1em} Other & 0.61 & (0.25, 1.16) & 1.30 & (0.50, 2.77)\\
{Single marital status} (baseline: With partner) & 1.40 & (1.06, 1.82) & 0.93 & (0.52, 1.51)\\
\multicolumn{5}{l}{{Smoking status} (baseline: Never)} \\
\hspace{1em} Former & 0.99 & (0.75, 1.28) & 1.18 & (0.67, 1.98)\\
\hspace{1em} Current & 1.02 & (0.60, 1.54) & 2.03 & (0.85, 4.25)\\
\multicolumn{5}{l}{{Insurance} (baseline: Government)} \\
\hspace{1em} Commercial & 0.71 & (0.49, 0.99) & 0.61 & (0.28, 1.15)\\
\hspace{1em} WCSC & 0.77 & (0.35, 1.39) & 0.99 & (0.34, 2.23)\\
\hspace{1em} Selfpay & 0.79 & (0.33, 1.50) & 0.90 & (0.28, 1.96)\\
\bottomrule
\end{tabular} \\
}
\end{table}
}

\begin{figure}[t!]
  \centering
  \includegraphics[width=1\textwidth]{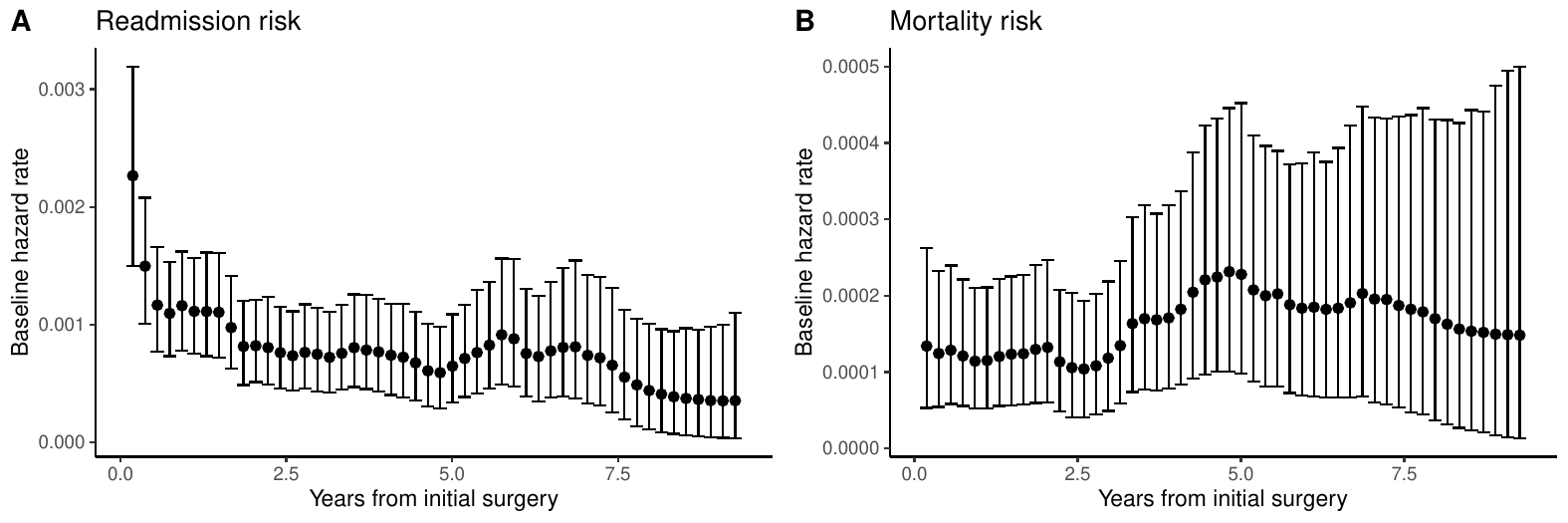}
  \caption{Posterior mean and 95\% credible interval for the baseline hazard rates. Panel A: Readmission risk. Panel B: Mortality risk.}
  \label{fig:res_bhr}
\end{figure}

Figure \ref{fig:res_bhr} shows the posterior means and 95\% credible intervals of the baseline hazard rates. For readmission risk, the estimated hazard closely mirrored the pattern observed in the exploratory analysis (Figure \ref{fig:EDA_hist_pop}), with a pronounced spike shortly after the initial surgical procedure. In contrast, for mortality risk, the model-fitted baseline hazard differed from the exploratory pattern in Figure \ref{fig:EDA_hist_pop}. After adjusting for age and other covariates, it exhibited an increasing trend, which better aligned with clinical expectations.

\begin{figure}[t!]
  \centering
  \includegraphics[width=1\textwidth]{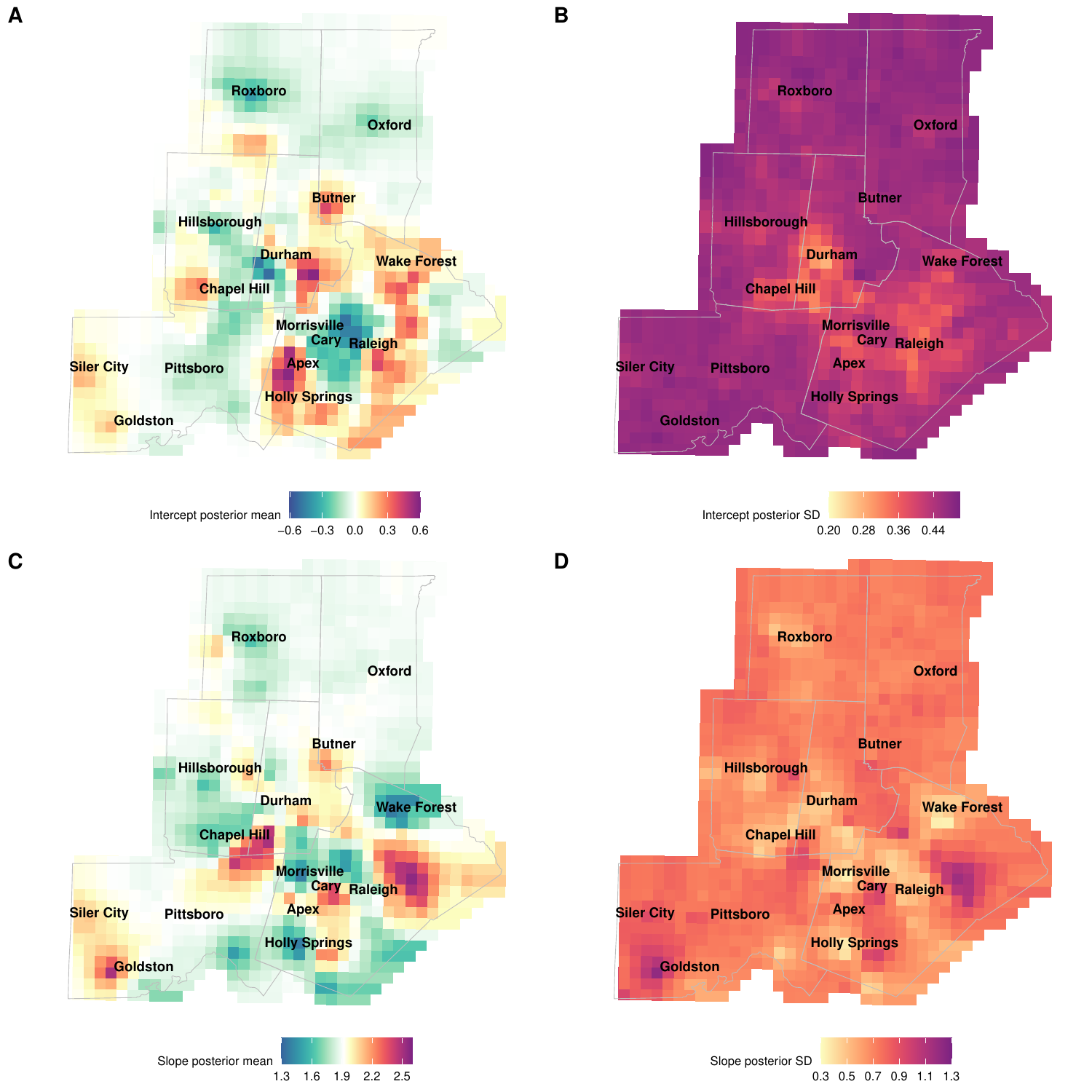}
  \caption{Posterior mean and standard deviation for readmission risk spatial surfaces. Panel A: Spatial intercept posterior mean. Panel B: Spatial intercept posterior standard deviation. Panel C: Spatial slope posterior mean. Panel D: Spatial slope posterior standard deviation.}\label{fig:res_r1}
\end{figure}

Figure \ref{fig:res_r1} shows model‐based spatial surfaces of readmission risk. Panel A displays the posterior mean of the spatial intercepts at all kriging locations, and panel B the corresponding posterior standard deviation. The spatial intercepts represent relative readmission risk as they were constrained to sum to zero over all kriging locations for identifiability. We observed patterns consistent with the exploratory analysis: Readmission risk is elevated in socioeconomically disadvantaged areas such as East Durham, and lower in regions with better healthcare resources, such as Chapel Hill and Cary. The posterior standard deviations are also sensible, with lower uncertainty in areas with more observations. Panel C presents the spatial slope surface for comorbidity on hazard ratio scale. For identifiability, we report combined fixed and spatial effects. Table \ref{tb:res_r1_beta} indicates that comorbidity is significantly and positively associated with readmission risk, and the spatial slope surface further reveals how this association varies across regions. Notably, in areas such as Wake Forest and Morrisville, the increase in readmission risk for higher comorbidity scores is less pronounced.

\begin{figure}[t!]
  \centering
  \includegraphics[width=1\textwidth]{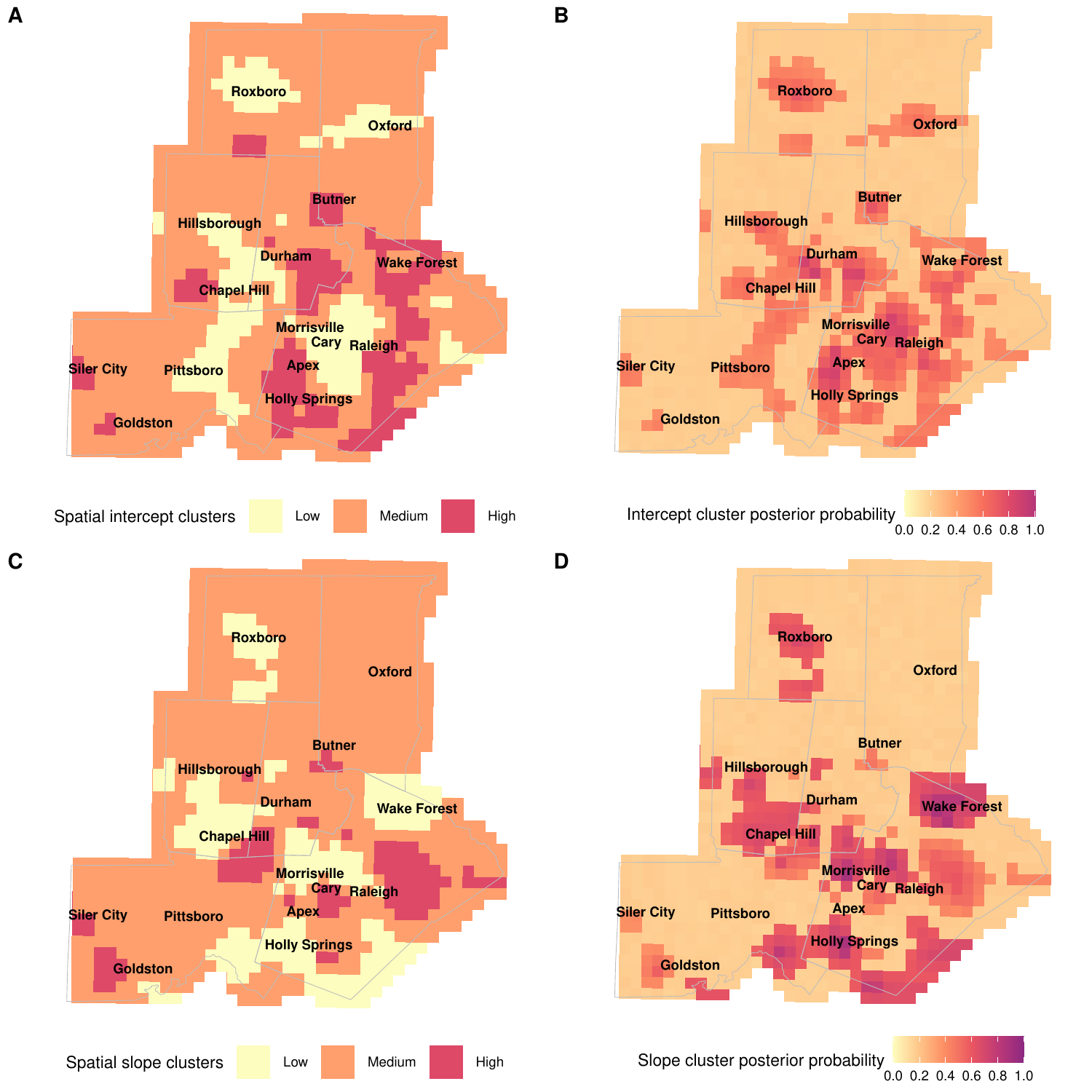}
  \caption{Clustering results for readmission risk spatial surfaces. Panel A: Spatial intercept clusters with cluster means: -0.17, 0.00, and 0.22. Panel C: Spatial slope clusters with cluster means on hazard ratio scale: 1.71, 1.91, and 2.23. Panel B and D: Posterior probability of being in the estimated intercept / slope cluster.}
  \label{fig:res_r1_clst}
\end{figure}

We further obtained Bayes optimal clusters under the K-means loss for the spatial surfaces. For practical reasons, we did not use cross-validation to select the number of clusters. Instead, we set $K=3$ as it provides spatial differentiation across the region, and at the same time gives clear, actionable interpretations for clinical researchers. Panels A and C of Figure \ref{fig:res_r1_clst} present clustering results for spatial intercepts and slopes respectively, while panels B and D report the associated cluster posterior probability. Higher posterior probability indicates greater certainty in the cluster assignment for a given location. For the spatial intercepts, for example, the results suggested with high posterior probability that areas such as Cary, Raleigh, Roxboro, and Oxford have relatively lower readmission risk, whereas East Durham, Apex, and Holly Springs exhibited higher risk. These findings highlighted regions that could be prioritized in policymaking, both to investigate factors contributing to lower risk, and to develop strategies to mitigate risk in the high-risk areas.

Overall, the proposed Bayesian competing risks spatial model was able to identify risk factors significantly associated with readmission risk while accounting for competing mortality risk and spatial confounders. It also provided insights on the spatial pattern of readmission risk, and highlighted high-risk areas that can inform downstream policy decisions. Results on mortality risk can be found in Section 6 of the Supplementary Materials.

\subsection{Sensitivity Analysis}\label{sec:sensi}

For the baseline hazard rates, we did not have prior information to gauge whether $k=50$ knots was sufficient to capture the true curves, and whether the hyperparameters were plausible. To ensure that our analysis results were robust to these hyperparameter choices, we conducted a sensitivity analysis to understand how different settings affect the model results. In particular, we considered a set of parameters which encourages lower autocorrelations in baseline hazard rates ($a_0=b_0=5$, $a_1=2$, $b_1=10$), a set of parameters which encourages higher autocorrelations ($a_0=b_0=0.1$, $a_1=2$, $b_1=80$), and a set with $k=100$ equally distanced knots instead of 50. The same analyses were carried out for these sensitivity tests, and the results remained similar in terms of model choice, significant risk factors, overall spatial patterns and clustering results of the estimated spatial surfaces. This showed that our analysis was robust to changes in the hyperparameters for the baseline hazard rates. See Section 7 of the Supplementary Materials for these results.

\section{Discussion}\label{sec:discussion}

In this manuscript, we introduced a Bayesian spatial competing risks framework that flexibly models point-referenced spatial random effects for time-to-event data with multiple risk types. Building on a Bayesian proportional cause-specific hazard model, we used independent Gaussian process (GP) priors to capture spatial dependence and implemented a Hilbert space GP (HSGP) approximation for computational scalability. Baseline hazard rates were modeled as piecewise constant, with a multiplicative gamma process prior that induces shrinkage and adaptively controls smoothness. A loss-based clustering method was applied on the spatial random effects to identify high-risk areas. This framework is broadly applicable to spatially structured competing risks data and, to our knowledge, represents the first integration of point-referenced spatial modeling into a competing risks proportional hazards model.

We evaluated the model through simulation and applied it to Duke EHR data to study hospital readmission among elderly patients with upper extremity fractures. The spatial competing risk model enabled inference on covariate effects while accounting for the competing mortality risk and unmeasured spatial confounders. And as shown by the simulation results, this framework produced more efficient inference than non-spatial models. Compared with areal spatial models, incorporating point-referenced spatial effects allowed finer spatial resolution of readmission risk. For example, panel A of Figure \ref{fig:res_r1} shows that while Apex is adjacent to Cary and Morrisville, their readmission risk profiles differ notably; the same applies to East and West Durham. Such granularity is often lost in areal models. These findings support prior evidence that patient outcomes reflect broader social determinants of health \citep{vrtikapa2025social}, and they can inform targeted clinical interventions. For example, providers treating patients in high-risk areas may schedule more frequent follow-ups or give more detailed discharge instructions. The resulting spatial clusters, combined with uncertainty quantification, also have policy implications: resources can be directed to high-certainty high-risk areas such as East Durham and Apex, while low-risk regions such as Cary and Morrisville can be studied to identify protective factors.

This application has a few limitations inherent to EHR data: (1) Duke EHR captures only patients who receive care within this health system, limiting generalizability, (2) readmissions to non-Duke hospitals are unobserved, leading to incomplete outcome data, (3) patient address data may be outdated or inaccurate, introducing noise to spatial modeling. In addition to the data limitations, we also noted several model limitations as discussed at length in the article: (1) the use of independent GP priors was unrealistically simple, (2) HSGP approximation tended to underestimate posterior uncertainty, (3) for the clustering results, we chose to report the posterior cluster probability after conditioning on the cluster centers which underestimates uncertainty. These compromises were made mainly for computational efficiency, and better interpretation.

There are several promising directions for future work. Extending the analysis to larger cohorts, such as patients with hip fractures, would enhance clinical relevance. Incorporating time-varying covariates from longitudinal encounters could potentially improve model efficiency. Finally, extending the model to allow time-varying coefficients could be valuable, as the association between covariates and the risk events may evolve over time.

\section{Supplementary Materials}

The Supplementary Materials contain inclusion and exclusion criteria for our patient cohort, demographics of the patient cohort, details on insurance type categories, proof of the kriging distribution under HSGP, additional results for the simulation study and EHR data application, and sensitivity analysis results. Simulation and real data analysis codes are available at \url{https://github.com/christineymshen/BCRSp}.

\bibliography{bibliography.bib}

@article{hesam2018spatial,
  title={{A spatial survival model in presence of competing risks for Iranian gastrointestinal cancer patients}},
  author={Hesam, Saeed and Mahmoudi, Mahmood and Foroushani, Abbas Rahimi and Yaseri, Mehdi and Mansournia, Mohammad Ali},
  journal={Asian Pacific Journal of Cancer Prevention},
  volume={19},
  number={10},
  pages={2947},
  year={2018}
}

@article{momenyan2022competing,
  title={{Competing risks model for clustered data based on the subdistribution hazards with spatial random effects}},
  author={Momenyan, Somayeh and Ahmadi, Farzane and Poorolajal, Jalal},
  journal={Journal of Applied Statistics},
  volume={49},
  number={7},
  pages={1802--1820},
  year={2022},
  publisher={Taylor \& Francis}
}

@article{riutort2023practical,
  title={{Practical Hilbert space approximate Bayesian Gaussian processes for probabilistic programming}},
  author={Riutort-Mayol, Gabriel and B{\"u}rkner, Paul-Christian and Andersen, Michael R and Solin, Arno and Vehtari, Aki},
  journal={Statistics and Computing},
  volume={33},
  number={1},
  pages={17},
  year={2023},
  publisher={Springer}
}

@article{solin2020hilbert,
  title={{Hilbert space methods for reduced-rank Gaussian process regression}},
  author={Solin, Arno and S{\"a}rkk{\"a}, Simo},
  journal={Statistics and Computing},
  volume={30},
  number={2},
  pages={419--446},
  year={2020},
  publisher={Springer}
}

@article{bhattacharya2011sparse,
  title={{Sparse Bayesian infinite factor models}},
  author={Bhattacharya, Anirban and Dunson, David B},
  journal={Biometrika},
  volume={98},
  number={2},
  pages={291--306},
  year={2011},
  publisher={Oxford University Press}
}

@book{banerjee2003hierarchical,
  title={{Hierarchical modeling and analysis for spatial data}},
  author={Banerjee, Sudipto and Carlin, Bradley P and Gelfand, Alan E},
  year={2003},
  publisher={Chapman and Hall/CRC}
}

@article{tian2023incidence,
  title={Incidence, causes, and risk factors of unplanned readmissions in elderly patients undergoing hip fracture surgery: an observational study},
  author={Tian, Miao and Wang, Zhijia and Zhu, Yanbin and Tian, Yunxu and Zhang, Kexin and Li, Xiuting},
  journal={Clinical Interventions in Aging},
  pages={317--326},
  year={2023},
  publisher={Taylor \& Francis}
}

@article{mathew2016risk,
  title={Risk factors for hospital re-presentation among older adults following fragility fractures: a systematic review and meta-analysis},
  author={Mathew, Saira A and Gane, Elise and Heesch, Kristiann C and McPhail, Steven M},
  journal={BMC Medicine},
  volume={14},
  number={1},
  pages={136},
  year={2016},
  publisher={Springer}
}

@article{lee2025risk,
  title={Risk factors for readmission within 30 Days after discharge following hip fracture surgery: a systematic review and meta-analysis},
  author={Lee, Kyung-Joo and Kim, Ji Wan and Kim, Chul-Ho},
  journal={Journal of Clinical Medicine},
  volume={14},
  number={8},
  pages={2779},
  year={2025},
  publisher={MDPI}
}

@article{liu2022heavy,
  title={{Heavy clinical and economic burden of osteoporotic fracture among elderly female Medicare beneficiaries}},
  author={Liu, J and Gong, T and Xu, X and Fox, KM and Oates, M and Gandra, SR},
  journal={Osteoporosis International},
  volume={33},
  number={2},
  pages={413--423},
  year={2022},
  publisher={Springer}
}

@techreport{HCUP_readmissions_2023,
  author       = {HCUP},
  title        = {Characteristics of 30-Day all-cause hospital readmissions, 2016--2020},
  institution  = {Agency for Healthcare Research and Quality},
  type         = {Statistical Brief},
  number       = {304},
  year         = {2023},
  url          = {https://hcup-us.ahrq.gov/reports/statbriefs/sb304-readmissions-2016-2020.jsp},
  note         = {Accessed: 2025-08-18}
}

@techreport{weiss2021readmissions,
  author       = {Weiss, Audrey J. and Jiang, H. Joanna},
  title        = {Overview of Clinical Conditions With Frequent and Costly Hospital Readmissions by Payer, 2018},
  institution  = {Agency for Healthcare Research and Quality (AHRQ)},
  year         = {2021},
  month        = {July},
  number       = {HCUP Statistical Brief \#278},
  address      = {Rockville, MD},
  url          = {https://www.hcup-us.ahrq.gov/reports/statbriefs/sb278-Conditions-Frequent-Readmissions-By-Payer-2018.pdf}
}

@article{wang2024national,
  title={National estimates of short-and longer-term hospital readmissions after major surgery among community-living older adults},
  author={Wang, Yi and Leo-Summers, Linda and Vander Wyk, Brent and Davis-Plourde, Kendra and Gill, Thomas M and Becher, Robert D},
  journal={JAMA Network Open},
  volume={7},
  number={2},
  pages={e240028},
  year={2024},
  publisher={American Medical Association}
}

@article{vrtikapa2025social,
  title={Social Determinants of Health: The Impact of This Overlooked Vital Sign},
  author={Vrtikapa, Katarina and Urmy, Farhana Hoque and Hoque, Farzana},
  journal={Journal of Brown Hospital Medicine},
  volume={4},
  number={3},
  pages={138072},
  year={2025}
}

@article{palmer2025quantifying,
  title={Quantifying sleep apnea heterogeneity using hierarchical Bayesian modeling},
  author={Palmer, Glenn and Dunson, David B},
  journal={arXiv preprint arXiv:2503.11599},
  year={2025}
}

@article{jiang2023pre,
  title={{Pre-pandemic assessment: a decade of progress in electronic health record adoption among US hospitals}},
  author={Jiang, John and Qi, Kangkang and Bai, Ge and Schulman, Kevin},
  journal={Health Affairs Scholar},
  volume={1},
  number={5},
  pages={qxad056},
  year={2023},
  publisher={Oxford University Press US}
}

@misc{ahrq2024elixhauser,
  title        = {Elixhauser Comorbidity Software Refined for ICD-10-CM: User Guide, Version 2024.1},
  author       = {{Agency for Healthcare Research and Quality}},
  organization = {Healthcare Cost and Utilization Project (HCUP)},
  year         = {2023},
  month        = {December},
  note         = {Accessed: 2024-06-16},
  url          = {https://hcup-us.ahrq.gov/toolssoftware/comorbidityicd10/comorbidity\_icd10.jsp}
}

@article{call2022factors,
  title={Factors associated with accurate reporting of public and private health insurance type},
  author={Call, Kathleen Thiede and Fertig, Angela R and Pascale, Joanne},
  journal={Health Services Research},
  volume={57},
  number={4},
  pages={930--943},
  year={2022},
  publisher={Wiley Online Library}
}

@Misc{stan2025,
    title = {{RStan}: the {R} interface to {Stan}},
    author = {{Stan Development Team}},
    note = {R package version 2.32.7},
    year = {2025},
    url = {https://mc-stan.org/},
  }

@article{hao2019role,
  title={The role of frailty in predicting mortality and readmission in older adults in acute care wards: a prospective study},
  author={Hao, Qiukui and Zhou, Lixing and Dong, Biao and Yang, Ming and Dong, Birong and Weil, Yuquan},
  journal={Scientific Reports},
  volume={9},
  number={1},
  pages={1207},
  year={2019},
  publisher={Nature Publishing Group UK London}
}

@article{bourriquen2024effect,
  title={Effect of frailty on unplanned readmission in older adults: A systematic review},
  author={Bourriquen, Maryline and Couderc, Anne-Laure and Bretelle, Fannie and Villani, Patrick},
  journal={Journal of Epidemiology and Population Health},
  volume={72},
  number={5},
  pages={202774},
  year={2024},
  publisher={Elsevier}
}

@misc{kff2024medicarecoverage,
  author       = {{Kaiser Family Foundation}},
  title        = {{A snapshot of sources of coverage among Medicare beneficiaries}},
  year         = {2024},
  howpublished = {\url{https://www.kff.org/medicare/a-snapshot-of-sources-of-coverage-among-medicare-beneficiaries/}},
  note         = {Accessed: 2025-09-19}
}

@misc{medpac2024databook,
  author       = {{Medicare Payment Advisory Commission}},
  title        = {{July 2024 data book: health care spending and the Medicare program}},
  year         = {2024},
  howpublished = {\url{https://www.medpac.gov/wp-content/uploads/2024/07/July2024_MedPAC_DataBook_Sec2_SEC.pdf}},
  note         = {Accessed: 2025-09-19}
}

@article{horvath2014modular,
  title={{Modular design, application architecture, and usage of a self-service model for enterprise data delivery: the Duke Enterprise Data Unified Content Explorer (DEDUCE)}},
  author={Horvath, Monica M and Rusincovitch, Shelley A and Brinson, Stephanie and Shang, Howard C and Evans, Steve and Ferranti, Jeffrey M},
  journal={Journal of Biomedical Informatics},
  volume={52},
  pages={231--242},
  year={2014},
  publisher={Elsevier}
}

@article{hurst2021development,
  title={Development of an electronic health records datamart to support clinical and population health research},
  author={Hurst, Jillian H and Liu, Yaxing and Maxson, Pamela J and Permar, Sallie R and Boulware, L Ebony and Goldstein, Benjamin A},
  journal={Journal of Clinical and Translational Science},
  volume={5},
  number={1},
  pages={e13},
  year={2021},
  publisher={Cambridge University Press}
}

@article{dignam2012use,
  title={The use and interpretation of competing risks regression models},
  author={Dignam, James J and Zhang, Qiang and Kocherginsky, Masha},
  journal={Clinical Cancer Research},
  volume={18},
  number={8},
  pages={2301--2308},
  year={2012},
  publisher={American Association for Cancer Research}
}

@article{haller2013applying,
  title={Applying competing risks regression models: an overview},
  author={Haller, Bernhard and Schmidt, Georg and Ulm, Kurt},
  journal={Lifetime Data Analysis},
  volume={19},
  number={1},
  pages={33--58},
  year={2013},
  publisher={Springer}
}

@Manual{survival-package,
    title = {A Package for Survival Analysis in R},
    author = {Terry M Therneau},
    year = {2024},
    note = {R package version 3.8-3},
    url = {https://CRAN.R-project.org/package=survival},
}

@article{wang2003bayesian,
  title={Bayesian analysis of bivariate competing risks models with covariates},
  author={Wang, Chen-Pin and Ghosh, Malay},
  journal={Journal of Statistical Planning and Inference},
  volume={115},
  number={2},
  pages={441--459},
  year={2003},
  publisher={Elsevier}
}

@article{samanta2021bayesian,
  title={Bayesian inference of a dependent competing risk data},
  author={Samanta, Debashis and Kundu, Debasis},
  journal={Journal of Statistical Computation and Simulation},
  volume={91},
  number={15},
  pages={3069--3086},
  year={2021},
  publisher={Taylor \& Francis}
}

@article{hu2009bayesian,
  title={{A Bayesian approach to joint analysis of longitudinal measurements and competing risks failure time data}},
  author={Hu, Wenhua and Li, Gang and Li, Ning},
  journal={Statistics in Medicine},
  volume={28},
  number={11},
  pages={1601--1619},
  year={2009},
  publisher={Wiley Online Library}
}

@book{ibrahim2001bayesian,
  title={Bayesian survival analysis},
  author={Ibrahim, Joseph G and Chen, Ming-Hui and Sinha, Debajyoti},
  year={2001},
  publisher={Springer Science \& Business Media}
}

@article{kalbfleisch1978non,
  title={{Non-parametric Bayesian analysis of survival time data}},
  author={Kalbfleisch, John D},
  journal={Journal of the Royal Statistical Society Series B: Statistical Methodology},
  volume={40},
  number={2},
  pages={214--221},
  year={1978},
  publisher={Wiley Online Library}
}

@article{hjort1990nonparametric,
  title={{Nonparametric Bayes estimators based on beta processes in models for life history data}},
  author={Hjort, Nils Lid},
  journal={The Annals of Statistics},
  pages={1259--1294},
  year={1990},
  publisher={JSTOR}
}

@article{watanabe2010asymptotic,
  title={{Asymptotic equivalence of Bayes cross validation and widely applicable information criterion in singular learning theory}},
  author={Watanabe, Sumio and Opper, Manfred},
  journal={Journal of machine learning research},
  volume={11},
  number={12},
  year={2010}
}

@Misc{looR,
    title = {{loo: Efficient leave-one-out cross-validation and WAIC for
      Bayesian models}},
    author = {Aki Vehtari and Jonah Gabry and Måns Magnusson and Yuling
      Yao and Paul-Christian Bürkner and Topi Paananen and Andrew
      Gelman},
    year = {2024},
    note = {R package version 2.8.0},
    url = {https://mc-stan.org/loo/},
  }

\end{document}